\documentclass[11pt,twocolumn,english]{article}
\usepackage{lmodern}

\usepackage[T1]{fontenc}
\usepackage[latin9]{inputenc}
\usepackage{geometry}
\usepackage{color}
\usepackage{babel}
\usepackage{array}
\usepackage{multirow}
\usepackage{amsmath}
\usepackage{amssymb}
\usepackage{graphicx}
\usepackage{rotfloat}
\usepackage[numbers,sort&compress]{natbib}
\usepackage[unicode=true,pdfusetitle,
 bookmarks=true,bookmarksnumbered=false,bookmarksopen=false,
 breaklinks=true,pdfborder={0 0 0},pdfborderstyle={},backref=false,colorlinks=true]
 {hyperref}

\makeatletter

\newcommand{\noun}[1]{\textsc{#1}}
\providecommand{\tabularnewline}{\\}

\@ifundefined{date}{}{\date{}}
\makeatother

\begin{document}
\title{Machine Learning in High Energy Physics: A review of heavy-flavor
jet tagging at the LHC}
\author{Spandan Mondal \thanks{Brown University, Providence, RI, USA.}{\addtocounter{footnote}{+2}}\and
Luca Mastrolorenzo \thanks{Ex-member, CMS Experiment, CERN.}}
\maketitle
\begin{abstract}
The application of machine learning (ML) in high energy physics (HEP),
specifically in heavy-flavor jet tagging at Large Hadron Collider
(LHC) experiments, has experienced remarkable growth and innovation
in the past decade. This review provides a detailed examination of
current and past ML techniques in this domain.

It starts by exploring various data representation methods and ML
architectures, encompassing traditional ML algorithms and advanced
deep learning techniques. Subsequent sections discuss specific instances
of successful ML applications in jet flavor tagging in the ATLAS and
CMS experiments at the LHC, ranging from basic fully-connected layers
to graph neural networks employing attention mechanisms. To systematically
categorize the advancements over the LHC's three runs, the paper classifies
jet tagging algorithms into three generations, each characterized
by specific data representation techniques and ML architectures. This
classification aims to provide an overview of the chronological evolution
in this field. Finally, a brief discussion about anticipated future
developments and potential research directions in the field is presented.
\end{abstract}
\global\long\def\pt{p_{\text{T}}}%
\global\long\def\bb{\text{b}\bar{\text{b}}}%
\global\long\def\cc{\text{c}\bar{\text{c}}}%

\section{Introduction}

The task of jet identification in hadron collider experiments has
seen steady development over the past decades. In particular, machine
learning (ML) has revolutionized the way we view and leverage various
observables associated with a jet and its constituents. A jet refers
to a collimated spray of particles that originate from a high-energy
collision. They may originate either from a single quark/gluon generated
in the hard scattering or from a hadronically-decaying Lorentz-boosted
heavy particle. While viewing the jet as a single object gives us
only limited information about the particle initiating the jet, looking
at its constituents provides us with additional information that can
be used to predict the type of the initiating particle. Several approaches
have been developed for this purpose over the past decades and these
approaches are generally referred to as jet tagging.

To analyze collision data in search for an interesting physics process
(signal), one usually chooses a particular jet reconstruction strategy,
along with a particular jet tagging algorithm (when necessary), depending
on the expected features of the final state. For example, if the process
involves the production of a hadronically-decaying heavy particle
(such as W, Z, or Higgs boson, or a top quark) with a large transverse
momentum ($\pt\gtrsim250$ GeV), one might use a so-called large-radius
``fat'' jet to reconstruct the entire decay within the jet radius.
Such jets are usually characterized by a multi-pronged structure.
Jet substructure techniques, and more recently ML-based classification
algorithms, have been developed to distinguish fat jets arising from
the aforementioned interesting heavy particle decays, from those arising
from other physics processes (backgrounds), such as a gluon splitting
into a pair of quarks. On the other hand, if the process generates
one or more quarks/gluons that are spatially isolated, the event can
be reconstructed using one or more small-radius ``thin'' jets. This
is usually followed by application of techniques to identify the ``flavor''
of the quark (up, down, strange, charm, bottom) or gluon that initiated
the jet. This approach of identifying jet flavor is referred to as
flavor tagging. The term flavor tagging can also be extended to specific
use cases involving fat jets, for example, in the context of identifying
heavy particles that decay into pairs of heavy-flavor quarks (bottom
quark-antiquark pair, $\bb$, or charm quark-antiquark pair, $\cc$).
For the rest of this paper, up, down, strange, charm, bottom quarks,
and gluons are denoted by u, d, s, c, b, and g, respectively, while
$\bb$ and $\cc$ are denoted by bb and cc, respectively. The term
``light-flavor'' is used to denote u, d, s, and g combined.

The challenge of flavor tagging has been tackled using several approaches
over the past three decades. These methods leverage certain observable
characteristics of heavy-flavor quark (b and c) hadronization and
subsequent decays that are absent in case of light-flavor quarks (u,
d, s) and gluons (g). As hadrons originating from b and c quarks (mostly
B and D hadrons, respectively, among others) have a sizable lifetime
(\textasciitilde 1.5 ps) \citep{PDG2022}, they usually traverse
a short distance within the detector systems (\textasciitilde few
millimeters to centimeters) before further decaying. This gives rise
to tracks that are displaced with respect to the primary point of
collision (primary vertex, PV) and hence appear to originate from
a displaced vertex (secondary vertex, SV). Methods leveraging the
presence and properties of displaced tracks and/or reconstructed SVs
were, therefore, developed to discriminate heavy-flavor jets from
light-flavor ones. Additionally, gluon-initiated jets usually have
a broader energy spectrum compared to quark-initiated jets, which
makes quark-gluon discrimination possible. In case of fat jets, jet
substructure methods leveraging predictions from the theory of quantum
chromodynamics (QCD) have been used for flavor tagging. A discussion
about early methods is presented in Sec. \ref{sec:Discussions}.

The advent of ML techniques, especially deep learning \citep{GoodBengCour16,Guest:2018yhq},
opened up a plethora of new possibilities of tackling the challenge
of jet flavor tagging. Unlike using high-level physics-motivated observables
as discussed above, ML-based methods allow us to build jet classification
algorithms by using information available from low-level observables
associated with jets. The general idea behind these approaches is
to train models to automatically leverage relevant information from
observables (input features) and the correlations between them, that
distinguish a heavy-flavor jet from a light-flavor one, using labeled
simulated samples of jets (supervised learning) or from unlabeled
collision data (unsupervised learning). However, these methods are
optimal only when physically-meaningful representations of jet constituents
are fed into corresponding artificial neural networks suited to process
such representations.

This article is structured as follows. Section \ref{sec:Machine-Learning-architectures}
discusses some general ML architectures and data representation methods
that have been used in the field. Section \ref{sec:Machine-Learning-in}
discusses some practical implementations of these in flavor tagging
at Large Hadron Collider (LHC) \citep{Evans:2008zzb} experiments.
In particular, applications in the CMS \citep{CMS:2008xjf} and ATLAS
\citep{ATLAS:2008xda} experiments at the LHC have been covered in
this review. Sections \ref{sec:Discussions} and \ref{sec:Conclusion}
present a discussion and a conclusion, respectively.

\section{ML architectures and data representation\label{sec:Machine-Learning-architectures}}

\subsection{Boosted Decision Trees\label{subsec:Boosted-Decision-Trees}}

Decision trees (DTs) \citep{hunt1966experiments,quinlan1983learning,quinlan1986induction}
can be viewed as models resembling flowcharts. Each node in a DT represents
a question and a decision based on a feature present in data. The
decision of each node, and hence the ``flow'' along nodes, are optimized
to lead to a correct prediction. Boosting \citep{FREUND1997119},
on the other hand, refers to an ensemble learning technique \citep{10.5555/648054.743935}
that sequentially updates the model based on the errors made by the
ensemble in the previous iteration. Boosted Decision Trees (BDTs),
therefore, refer to models structured as DTs that leverage the technique
of boosting to iteratively refine predictions based on the training
dataset to perform classification and regression tasks \citep{Breiman:1984aa}.

BDTs are suitable for use with tabular data, represented as columns
and rows of numbers and categorical variables. Each column of the
table usually represents a feature and each row represents a unit
(e.g., a jet). In supervised learning approaches, the correct (target)
values are provided as additional columns in the training dataset.

Gradient Boosting Machines (GBMs) \citep{10.1214/aos/1013203451}
introduce a more nuanced approach towards optimization of BDTs by
making use of an technique called gradient descent. This introduces
a \textit{loss function} that quantifies the difference between the
predictions and the target values in the training dataset using a
predefined function. The gradient descent technique minimizes the
loss function by updating the model parameters in the direction opposite
to the gradient of the loss function. This allows fine-tuning of the
model parameters and a more efficient convergence to the optimal solution.
Gradient descent also allows the model to learn complex non-linear
patterns in the input phase space.

Traditional BDTs as well as GBMs are widely used in a range of disciplines
including Physics, in tasks involving simple, tabular data. Popular
packages involve \noun{Scikit-Learn \citep{scikit-learn}} (Python
\citep{10.5555/1593511}), \noun{XGBoost} \citep{Chen_2016} (Extreme
Gradient Boosting), and TMVA \citep{Hocker:1019880} (ROOT \citep{Brun:1997pa}).

\subsection{Dense Neural Networks\label{subsec:Dense-Neural-Networks}}

The simplest form of Neural Networks (NNs) \citep{rosenblatt1957perceptron,hopfield1982neural}
consist of an input layer, a single hidden layer, and an output layer.
Each node in a layer, referred to as a neuron, is densely connected
to neurons in the subsequent layer, making a ``fully-connected''
structure. The flow of information is one-directional, from the input
layer through the hidden layer to the output layer, which is why these
architectures are also referred to as feed-forward NNs.

The connections between layers have associated weights, that are determined
through the training process. Predefined activation functions \citep{2021arXiv210914545D}
determine the output of each neuron and introduce non-linearity which
helps the network capture complex patterns and relationships in data.
They determine the activation of a neuron based on the weights of
its inputs. 

Feed-forward Deep NNs (DNNs) \citep{Fukushima:1980aa} extend the
architecture of simple feed-forward NNs by introducing additional
hidden layers in between the input and output layers. This additional
``depth'' allows the network to automatically extract and leverage
hierarchical features, intricate relationships, and non-linearities
in the training data. Each layer learns progressively more abstract
features and complex patterns which makes them suitable for large
and higher-dimensional datasets.

Feed-forward NNs and DNNs are suitable for use with tabular data similar
to BDTs. However, DNNs are usually capable of handling larger datasets
and a larger number of input features as they can effectively capture
complex, non-linear patterns in data. In this paper, the term ``shallow
NN'' is used to describe dense NNs with one or two hidden layers,
each with a small number (\textasciitilde 10) of nodes. On the other
hand, ``DNN'' is used to refer to NNs with a larger number of hidden
layers and architectures that are not necessarily dense, feed-forward,
or fully-connected.

The term Multilayer Perceptron (MLP) is also used to describe dense
NNs. The simplest kind of MLP, a feed-forward MLP with one hidden
layer, is essentially the same as a shallow NN. On the other hand,
MLPs with additional hidden layers, which still have a feed-forward
structure, resemble dense DNNs. In this paper, the terms ``dense'',
``fully-connected'', ``feed-forward'', and ``MLP'' are used
interchangeably to describe this class of NNs.

\subsection{Convolutional Neural Networks}

Convolutional Neural Networks (CNNs) have revolutionized the field
of image recognition and computer vision \citep{Krizhevsky2012ImageNetCW}.
The architecture of CNNs is inspired by the visual processing hierarchy
of the human brain, where initial layers in the hierarchy capture
simple features like edges and textures, while deeper layers capture
more complex patterns like composite objects and structures. 

CNNs make use of convolutional layers, where filters (or kernels)
slide across the input data and capture local patterns. In the context
of a large image expressed as rectangular pixellated data, one can
imagine the filter as a rectangle with a smaller area, systematically
sliding across the image and producing an output corresponding to
each spatial position. The output for each position depends on the
local information of the image (pixels in the kernel's receptive field)
and the values of the kernel itself. The sliding process thus generates
a new feature map typically smaller than the original image. The convolutional
filters are adapted iteratively in the CNN training process such that
they autonomously detect and learn representations that are most relevant
to the task at hand. The associated weights are shared across all
neurons in a particular feature map.

Convolutional layers are usually followed by pooling layers. The pooling
operation involves sliding a rectangular pooling window in a similar
way across the generated feature map. The main goal of the pooling
layer is to reduce the spatial dimension by selecting representative
values within local neighborhood of the feature maps. The representative
values are usually evaluated through simple mathematical operations
like taking the maximum of or averaging the pixel values of the feature
map in the window's receptive field. This not only helps in down-sampling
the feature maps but also abstracts information by retaining only
the relevant features from local regions of the image.

Unlike fully-connected NNs, neurons in each layer are connected to
a neurons in a small spatial region of the previous layer. This lets
the network capture local patterns and leverage the fact that adjoining
pixels may contain related information without being sensitive to
the exact spatial position of the feature relative to the full image.
Again unlike fully-connected NNs, CNNs make use of weight sharing,
which essentially means the same filter is applied to the entire image,
and therefore use fewer trainable parameters.

CNNs are well-suited for tasks involving image recognition and classification.
Images fed as input to CNNs are generally represented as pixels with
one or multiple channels. In the context of jet tagging, jets can
be viewed as images on the detector surface with particle energy deposits
acting as intensities of the pixels of the image.

\subsection{Recurrent Neural Networks\label{subsec:Recurrent-Neural-Networks}}

Recurrent Neural Networks (RNNs) \citep{hopfield1982neural,rumelhart1985learning,Graves2012}
are designed for sequential data and time series. RNNs process data
sequentially such that the output at each step is influenced by a
hidden state that encapsulates information from previous temporal
steps. This lets the RNN maintain ``memories'' of previous inputs
and learn temporal dependencies.

Long Short-Term Memory networks (LSTMs) \citep{hochreiter1997long}
are specialized RNNs that make use of memory cells for a better handling
of sequential data. These memory cells store and manage information
over extended sequences, by selectively storing, discarding, and outputting
information. This allows LSTMs to capture long-term dependencies in
sequential data.

Gated Recurrent Units (GRUs) \citep{2014arXiv1406.1078C,2014arXiv1412.3555C},
another variant of RNNs, offer a streamlined architecture without
using memory cells but employ similar gating mechanisms to capture
dependencies in sequential data. GRUs use update gates and reset gates
to control the flow of information, making them computationally efficient
and effective for tasks involving sequential information.

RNNs and LSTMs have been widely used in Natural Language Processing
(NLP) \citep{Mikolov2010RecurrentNN}, time-series analysis \citep{279188,2016arXiv160601865C},
machine translation \citep{2014arXiv1406.1078C,2015arXiv150804025L},
speech processing \citep{2013arXiv1303.5778G}, and other domains
where interpreting information across multiple steps is crucial. An
interesting feature of LSTMs is their ability to handle sequences
of varying lengths. This is particularly relevant in the field of
jet tagging since jets naturally contain a varying number of clustered
particles. Additionally, the ability of LSTMs to selectively store
and forget information allows them to create meaningful summaries
or averages over information (particles) available in a sequence (jet).
This helps LSTMs identify, for example, particles in a jet that contribute
the most to jet identification. 

The utility of RNNs extends naturally to structured data like trees.
Each node in a tree represented as a step in the sequence allows RNNs
to capture dependencies and relationships within the hierarchical
structure of the tree. As jets are clustered sequentially from individual
particles, their clustering history can be represented as binary trees.
Operating on trees, RNNs can take multiple inputs per level in the
tree to add to the internal state at each step.

In the context of preprocessing data to be sequential, \textit{tokenization}
is a crucial step. It involves breaking down raw inputs into smaller
units called tokens. Each token represents a discrete unit, facilitating
the representation of the sequential information in a format suitable
for RNNs and LSTMs.

\subsection{Transformers\label{subsec:Transformers}}

A crucial feature of transformer models is the self-attention mechanism
\citep{NIPS2017_3f5ee243,2014arXiv1409.0473B}, that allows the model
to assign importance (weights) to different parts of the input sequence
differently. This allows them to capture long-range dependencies,
extending the concept from LSTMs. Multi-head Attention (MHA) extends
the concept by employing multiple sets of parameters (heads) to independently
compute attention scores in parallel. The outputs from all the heads
are then combined linearly with learnable weights. This allows the
model to capture various patterns and dependencies within the input
sequences simultaneously with different choice of attention parameters,
making the model more versatile and robust.

Unlike RNNs, transformers can operate in parallel while computing
attention weights. Furthermore, each layer in transformer models can
operate independently of the others. Parallelization makes transformer
models highly efficient, especially when training with large datasets.
It also contributes to faster training and prediction times in general. 

Transformers have revolutionized the field of NLP, achieving unprecedented
performances in tasks such as language understanding, translation,
and question-answering. The concept of attention has also been applied
to a wide range of tasks in ML as an effective tool for capturing
important information selectively. Transformers also excel at classification
tasks, which makes them relevant in jet flavor tagging. In classification
tasks, a specialized token referred to as a class token is positioned
at the start of an input sequence during tokenization. The hidden
state associated with this class token is then updated throughout
the sequence and utilized for making predictions through a dedicated
classification head.

\subsection{Deep Sets\label{subsec:DeepSets}}

In contrast to tabular data in dense NNs, grids in CNNs, and ordered
sequences in RNNs, physical jets do not intrinsically have an ordered
representation. Orders are usually imposed (for example, by sorting
jet constituents in descending order of their $\pt$) to make their
representations suitable as inputs for the aforementioned networks.
However, the choice of ordering is not unique and a more natural representation
of a jet is as an unordered set of particles.

Deep Sets \citep{2017arXiv170306114Z} provide a framework to handle
data that can be represented as sets of unordered units, where the
relationships between elements in the set matter while their orders
are inconsequential. The key component in the Deep Set implementation
is the approximation of a set function: a function that takes a set
of elements as input and produces fixed-size representation of the
entire set. The function is required to be permutation invariant to
the order of objects in the set. The architecture acts on each instance
in the input set and transforms each element, possibly through several
layers, into some abstract representation. The representations from
the entire set are added up and the output is processed using dense
NNs through non-linear transformations. The set function is usually
designed in a way to make use of weight sharing, which ensures that
the network learns consistent transformers across all elements and
results in fewer learnable parameters.

Thus Deep Sets can not only handle sequences of varying lengths but
also provide consistent representations of sets with different ordering
of elements, which is a more natural choice for representing jets.
This improves the generalization capabilities of the model without
relying on specific element orders. Permutation invariance also simplifies
the training process and training optimization and leads to faster
and stable convergence. 

\subsection{Graph Neural Networks\label{subsec:Graph-Neural-Networks}}

Graph Neural Networks (GNNs) \citep{2016arXiv160609375D,Bronstein_2017,2018arXiv180601261B}
are tailored to handle data with well-defined and complex relationships.
They are designed to process graphs: data structures that consist
of nodes that represent entities, interconnected by edges that represent
relationships and interactions between the different entities. They
are, therefore, the natural choice for data structures characterized
by interconnected and interdependent nodes.

A key feature of GNNs is the concept of \textit{message passing}.
Information is propagated between nodes connected by an edge. Each
node aggregates information from its connected neighbors in a local
as well as a global context. This allows GNNs to learn patterns of
information flow, identify nodes important to the task at hand, and
leverage abstract relationships between all nodes in the network.

In the field of HEP, GNNs are often used to model particles and/or
physics objects represented as nodes of a graph, and pairwise features
between particles as edges. Such representations are referred to as
point clouds in general, and sometimes as particle clouds in HEP.

\subsection{Dynamic Graph Convolutional Neural Networks \label{subsec:Dynamic-Graph-Convolutional}}

Dynamic Graph Convolutional Neural Networks (DGCNN) \citep{10.1145/3326362}
represent an evolution in the realm of GNNs by extending their capabilities
to characterize local features in the global structure of the data.
DGCNNs are equipped to handle graphs that are dynamically computed
in each layer of the network starting with the input graph representation
at the first layer, as opposed to GNNs that act on a static input
graph. This is achieved using a stackable NN module referred to as
edge convolution (EdgeConv). In the context of HEP, sequential EdgeConv
operations, starting with original particle coordinates and features,
lets one create several layers of graph representations of particle
clouds. This is similar to processing images with CNNs where one can
create hierarchical (increasingly abstract) representations of an
image by sequentially applying convolutional filters starting with
the original image.

The EdgeConv operation begins with a graph whose nodes represent particles,
and edges for each node represent the difference in feature vectors
of the $k$ nearest neighboring nodes from the central node. An edge
function with learnable parameters is then used to project each point
into a latent feature space, where the point is effectively assigned
new coordinates. Thus, a single EdgeConv operation transforms one
point cloud into another point cloud with the same number of points,
but with different feature vectors for each point. Repeated EdgeConv
operations can be applied to generate hierarchical representations
such that feature space structures in the deeper layers can group
related points together regardless of their spatial separation in
the original graph structure. 

Similar to convolution operations in CNNs, EdgeConv operations incorporate
local neighborhood information and can be stacked to learn global
spatial properties. In the context of HEP, this is helpful in capturing
characteristics and relationships between particles that are spatially
well-separated in the original representation on the detector surface. 

\subsection{Summary}

This section describes a few ML architectures that are widely used
and especially relevant to the task of jet tagging in HEP. Different
architectures are suitable for different kinds of data and their representations.
A brief, non-comprehensive overview of architectures and corresponding
datasets is presented in Table \ref{tab:Archs}. 

\begin{table}
\begin{centering}
\renewcommand{\arraystretch}{1.5}
\begin{tabular}{cc}
\hline 
\textbf{Architecture} & \textbf{Suitable for}\tabularnewline
\hline 
BDT & Tabular/structured data\tabularnewline
Dense NN/MLP & Tabular/structured data\tabularnewline
CNN & Images, grids\tabularnewline
RNN/LSTM & Sequences, trees, text\tabularnewline
Transformer & Sequences, text, large datasets\tabularnewline
Deep Set & Unordered sets, point clouds\tabularnewline
GNN & Graphs, point clouds\tabularnewline
DGCNN & Graphs, point clouds\tabularnewline
\hline 
\end{tabular}
\par\end{centering}
\caption{\label{tab:Archs}A few common ML architectures (first column) and
non-comprehensive lists of data representations (second column) they
are suited for. }

\end{table}

\section{ML in heavy-flavor jet identification at hadron colliders\label{sec:Machine-Learning-in}}

\begin{figure*}
\begin{centering}
\includegraphics[width=1\textwidth]{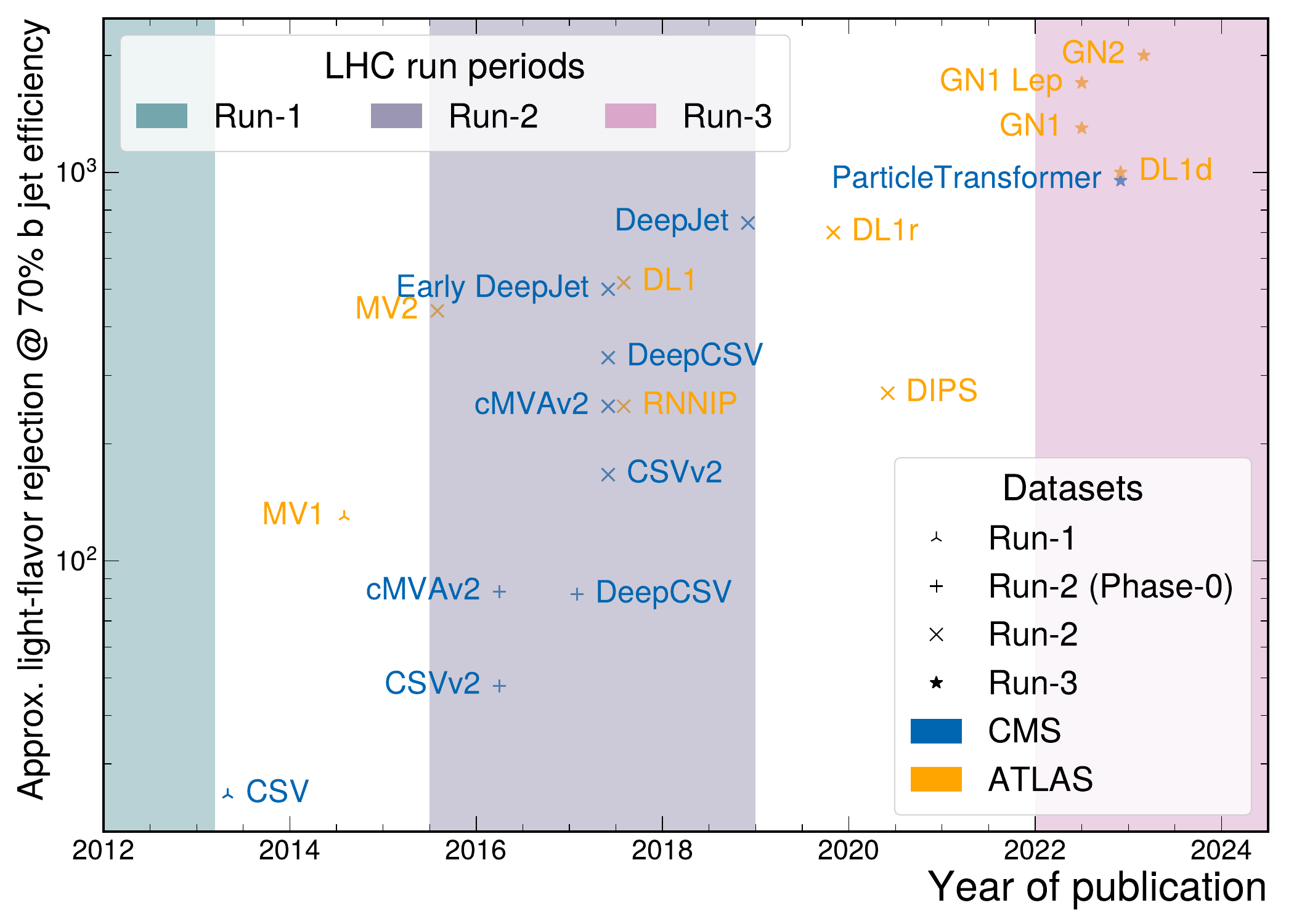}
\par\end{centering}
\caption{\label{fig:Approximate-light-flavor-jet}Approximate light-flavor
jet rejection rates at a b jet efficiency of 70\% (y-axis) achieved
by different ML-based single-pronged jet flavor tagging algorithms
used in the CMS and ATLAS experiments. The x-axis shows the approximate
years the algorithms were published by the respective experiments.
The vertical color bands indicate the timespan of the runs at the
LHC, while the different markers indicate different datasets reconstructed
with different run conditions on which the taggers were trained and/or
evaluated. Run-2 data reconstruction in CMS is split into two parts
corresponding to before (Phase-0) and after the CMS Phase-1 pixel
detector upgrade \citep{CMSTrackerGroup:2020edz}. It is important
to note that (a) the CMS and ATLAS detectors, along with their subsystems,
possess different detection and reconstruction capabilities, (b) there
are differences in jet reconstruction techniques and quality criteria
between the two experiments, (c) these techniques and criteria evolve
over time, and (d) definitions of tagger scores, and hence the degree
of discrimination specifically against light-flavor jets, vary between
experiments (e.g. using tagger output scores versus employing transformations
like that in eq. \ref{eq:DL1b}). Consequently, directly comparing
tagger performance across experiments may not accurately reflect algorithm
capabilities and must be interpreted carefully.}
\end{figure*}

Hadron collider experiments have adopted several ML-based strategies
for jet tagging over the years. These strategies can be broadly classified
into single-pronged jet tagging typically used in conjunction with
thin jets, and multi-pronged jet tagging used with fat jets. The following
subsections discuss these two categories separately. In this paper,
flavor tagging algorithms are further classified into three generations.
Taggers developed with BDTs and shallow NNs during LHC Run-1 \citep{Alemany-Fernandez:1631030}
and early Run-2 \citep{Wenninger:2668326}, are included in the first
generation of taggers. In the second generation, taggers using early
DNNs, CNNs, and RNNs, developed during Run-2 and implemented as standard
taggers by the end of Run-2 data reconstruction, are included. Finally,
more advanced taggers developed after Run-2 and commissioned with
early Run-3 \citep{Fartoukh:2790409} data are included within the
third generation of taggers. These include more complex algorithms
implementing Deep Sets, GNNs, DGCNNs, and transformer architectures,
relying on representing jets as unordered sets and particle clouds.
A summary of the taggers is presented in Table \ref{tab:Flavor-tagging-algorithms}
at the end of this section. Additionally, a plot depicting the evolution
of single-pronged jet tagging performance over the last decade is
presented in Fig. \ref{fig:Approximate-light-flavor-jet}. 

\subsection{Single-pronged jet tagging}

Individual quarks and gluons usually give rise to small-radius ``thin''
jets. In both CMS and ATLAS experiments, these are clustered using
the anti-$k_{\text{T}}$ (AK) \citep{Cacciari:2008aa,Cacciari:2011ma}
algorithm. The distance parameter $\Delta R$ used by the experiments
for thin jets was 0.5 or 0.4 during Run-1, 0.4 during Run-2, and 0.4
in Run-3. These jets are referred to as AK5 or AK4 jets, depending
on $\Delta R$.

\subsubsection{First generation: BDTs and shallow NNs\label{subsec:First-generation:-BDTs}}

The Combined Secondary Vertex (CSV) algorithm \citep{CMS:2012feb}
was developed in CMS during Run-1 of the LHC. This approach combines
a small set of high-level physics-motivated variables, that are expected
to have different distributions for b, c, and light-flavor jets, such
as vertex kinematics, track kinematics, multiplicities, etc. The number
of variables ranges between 2 to 8 depending on whether an SV is reconstructed
\citep{Waltenberger_2007,Waltenberger:1166320,CMS:2011yuk} within
the jet. Two likelihood ratios are built from these variables, to
discriminate b from light-flavor, and b from c jets, respectively.

In Run-2 of the LHC, the CSV algorithm was updated by implementing
a feed-forward MLP (cf. Sec. \ref{subsec:Dense-Neural-Networks})
with one hidden layer. The new NN-based approach was dubbed as CSVv2
\citep{CMS-PAS-BTV-15-001,Sirunyan:2017ezt}. Even though a likelihood-based
method with a small number of physics-motivated variables is a simple
and physically-interpretable approach, the update from using likelihood
ratios in the CSV algorithm in Run-1 to using an MLP in CSVv2 in Run-2
allowed the incorporation of additional input variables in the input
layer of the MLP. Thus, up to 22 variables, including information
of up to 4 tracks, additional track and SV variables, and overall
kinematic features of the jet, were used as inputs to the algorithm.
The number of nodes in the hidden layer was set to twice the number
of input variables for effective projection and extraction of features
from the input data. As a result of additional input information and
the updated architecture, the efficiency of selecting b jets at approximately
10\% light-jet mistag rate went up from around 74\% in CSV to around
82\% in CSVv2. More detailed efficiency values at different mistag
rates can be found in Refs. \citep{CMS-DP-2015-056,CMS-DP-2017-012}.

In addition to track and SV variables, soft-leptons arising from semileptonic
decays of b and c hadrons can also provide handles to discriminate
heavy-flavor jets from light-flavor ones. Such jets, however, account
for only about 20\% (10\%) of b (c) hadron decays. In Run-2 of CMS,
BDT-based (cf. Sec. \ref{subsec:Boosted-Decision-Trees}) soft-lepton
(SL) taggers \citep{Sirunyan:2017ezt} were used as jet flavor taggers.
These algorithms, called soft-electron (SE) and soft-muon (SM) taggers,
make use of 2D and 3D impact parameter (IP) significances\footnote{IP significance is defined as the ratio of the IP value to its uncertainty.}
of the lepton, the angular distance between the jet axis and the lepton,
the ratio of the $\pt$ of the lepton to that of the jet, the $\pt$
of the lepton relative to the jet axis, and an MVA-based electron
quality (in case of SE only) as inputs. 

In Run-2, the CMS experiment also implemented a combined tagger named
combined MultiVariate Analysis version 2 (cMVAv2) \citep{Sirunyan:2017ezt}.
This tagger combined the outputs of six other heavy-flavor taggers,
including both physics-motivated taggers and ML-based taggers like
CSVv2, SE, and SM, using a GBM model (cf. Sec. \ref{subsec:Boosted-Decision-Trees})
as a BDT. The cMVAv2 algorithm achieves a gain of about 3--4\% in
b jet identification compared to the CSVv2 algorithm, indicating the
importance of additional input information. 

Dedicated c jet tagging algorithms were also developed during Run-2
of CMS. This development was motivated by both the availability of
more advanced ML technologies in Run-2, and the general switch of
the focus of the physics program at the LHC experiments from discovery
of the Higgs boson \citep{ATLAS:2012yve,CMS:2012qbp} in Run-1 to
characterization of the nature of the Higgs boson \citep{CMS:2022dwd}
in Run-2. As properties of c jets lie somewhere in between that of
b and light-flavor jets, two different taggers to discriminate c from
b (CvsB), and c from light-flavor (CvsL) jets were developed \citep{Sirunyan:2017ezt,CMS-PAS-BTV-16-001}.
These taggers combined the same kind of track, SV, and soft-lepton
information used in CSVv2 and SL taggers. Two separate GBM models
as BDTs were trained to implement the CvsB and CvsL taggers. Physics
analyses in CMS involving c jets in the final state implemented selection
criteria on both the CvsL and CvsB output scores to select c jets
while rejecting both light-flavor and b jets at certain misidentification
rates.

The ATLAS experiment used likelihood-based approaches exploiting IP
and SV information separately to construct low-level b tagging algorithms
in Run-1 \citep{ATLAS:2015thz,ATL-PHYS-PUB-2017-011}. Additionally,
the JetFitter algorithm \citep{ATLAS:2009zsq,ATL-PHYS-PUB-2018-025}
based on a shallow NN was also developed. JetFitter was designed to
exploit the topological structure of b and c hadron decays inside
jets. It takes 8 variables as input, which include vertex and track
multiplicities, the vertex mass, energy fraction carried by the tracks,
vertex flight distance significance, jet $\pt$, and jet pseudorapidity.
The JetFitter architecture consisted of two hidden layers, with 12
and 7 nodes, respectively. The output layer consisted of 3 nodes,
representing the probabilities of the jet being a b, c, or light-flavor
jet, respectively. 

In addition to low-level tagging algorithms, the ATLAS experiment
in Run-1 used a MLP-based combined tagging algorithm called MV1 \citep{ATLAS-CONF-2014-046,ATLAS:2015thz}.
The MV1 algorithm was designed to combine the outputs of the two low-level
likelihood-based taggers exploiting IP and SV information. As an NN,
the MV1 algorithm had the advantage of automatically leveraging complex
correlations between its inputs. The architecture consisted of 2 hidden
layers with 3 and 2 nodes, respectively. The output layer consisted
of only 1 node indicating the final discriminant value. The MV1 algorithm
improved the light-flavor jet background rejection by a factor of
\textasciitilde 2 compared to the JetFitter algorithm, at a b jet
tagging efficiency of 70\%. The improvement can be attributed to MV1's
ability to combine IP and SV information as opposed to the SV-based
JetFitter algorithm.

In early Run-2 of the LHC, the ATLAS experiment implemented the MV2
algorithm \citep{ATL-PHYS-PUB-2015-022,ATL-PHYS-PUB-2015-039,ATL-PHYS-PUB-2017-013}.
As opposed to the NN-based MV1, the MV2 is built on a BDT-based architecture.
The key distinction between MV1 and MV2 lies in the handling of input
IP and SV information. MV1 utilizes the outputs of intermediate low-level
tagging algorithms, whereas MV2 directly takes as input the variables
employed as inputs to the low-level taggers. Additionally, the MV2
algorithm makes use of the inputs to the JetFitter algorithm. This
results in a total of 24 input variables to the BDT-based architecture.
This streamlined approach not only simplifies the overall flavor tagging
process but also enables MV2 to directly leverage correlations among
the lower-level inputs, avoiding the inevitable loss of information
that occurs when condensing input information into just two or three
numbers. 

Several versions of the MV2 algorithm were trained, varying the proportions
of light-flavor and c jets used as backgrounds in the training, with
the aim of optimizing the respective networks to strike a balance
between light-flavor jet and c jet discrimination. Algorithms dubbed
MV2c00, MV2c10, and MV2c20 use only light-flavor jets, an admixture
of 90\% light-flavor and 10\% c jets, and an admixture of 80\% light-flavor
and 20\% c jets, respectively, as backgrounds. As expected, the MV2c20
tagger significantly improves c jet rejection (\textasciitilde 4
times better at 50\% b efficiency) with only a slight degradation
(factor of \textasciitilde 1.5 at 50\% b efficiency) in light-flavor
jet rejection, compared to the MV2c00 tagger. The MV2c10 and MV2c20
variants were, however, trained with altered c jet fractions in 2016
while retaining the nomenclatures. The details about the optimization
can be found in Ref. \citep{ATL-PHYS-PUB-2016-012}. The intermediate
MV2c10 tagger was used as the standard tagger in ATLAS in Run-2 \citep{ATLAS:2018sgt,ATLAS-CONF-2018-045,ATLAS-CONF-2018-006}.
When comparing the Run-1 MV1 and Run-2 MV2 algorithms, MV2 improves
light-jet rejection by a factor of \textasciitilde 4 and c jet rejection
by a factor of 1.5--2 for a b jet efficiency of 70\%.

\subsubsection{Second generation: Early DNNs\label{subsec:Second-generation:-Early}}

The advent of deep learning led to the development of the DeepCSV
algorithm \citep{CMS-DP-2017-005,Sirunyan:2017ezt} in the CMS experiment.
This is an update to the CSVv2 algorithm and incorporates more hidden
layers, more nodes per layer, and the kinematics of up to six tracks
as inputs. The total number of inputs to DeepCSV can be up to 66 per
jet. The architectures consists of four hidden layers and 100 nodes
per layer, while maintaining the fully-connected and feed-forward
structure. Furthermore, the architecture is implemented as a multiclassifier,
which means the output layer has 5 nodes representing 5 types of jets,
namely b, bb, c, cc and light-flavor. As opposed to\noun{ Scikit-Learn}
used for BDT and shallow NN training, these trainings made use of
the \noun{Keras} \citep{chollet2015keras} and \noun{TensorFlow \citep{tensorflow2015-whitepaper}}
libraries. 

The DeepCSV algorithm achieves a 3--9\% improvement in b jet efficiency
compared to CSVv2 depending on the selection criterion used on the
output score. The performance is similar to, or on par with cMVAv2
at low purity (high light-jet mistag) regions, while the performance
is significantly superior to cMVAv2 at high purity selections. The
improved performance of DeepCSV without using explicit soft-lepton
information in jets can be attributed to the use of a deeper NN capable
of taking a vastly larger number jet variables as inputs. 

Owing to the multiclassifier nature of the DeepCSV algorithm, it is
suitable for c jet tagging as well. The output scores can be transformed
to define c tagging scores that perform the same task as dedicated
CvsB and CvsL BDT trainings, as follows:
\begin{align}
\text{DeepCSV CvsB} & =\frac{P(\text{c})+P(\text{cc})}{1-P(\text{udsg})},\nonumber \\
\text{DeepCSV CvsL} & =\frac{P(\text{c})+P(\text{cc})}{1-\left(P(\text{b})+P(\text{bb})\right)}.\label{eq:DeepCSV}
\end{align}
Charm jet tagging with DeepCSV also achieves slightly better performance
than with dedicated BDT trainings.

The first algorithm based on deep learning implemented in the ATLAS
experiment was the DL1 algorithm \citep{ATL-PHYS-PUB-2017-013,ATLAS:2019bwq}.
The variables used as inputs to DL1 are the same as the MV2 algorithm
(cf. Sec. \ref{subsec:First-generation:-BDTs}) with the addition
of so-called ``JetFitter c tagging variables''. The latter include
properties of a tertiary vertex, consistent with the hypothesis of
a b hadron further decaying to a c hadron, such as flight distance,
vertex mass and number of tracks, energy, energy fraction, and rapidity
of the tracks associated with the secondary and tertiary vertices.
This results in a total of 28 input variables. The DL1 architecture
consists of 8 fully-connected hidden layers, each with a decreasing
number of nodes ranging from 78 to 6, and 3 Maxout layers \citep{2013arXiv1302.4389G}.
Similar to DeepCSV, DL1 is also a multiclassifier with dedicated output
nodes corresponding to probabilities of the jet being a b, c, or light-flavor
jet. The b tagging discriminator used in the ATLAS experiment is defined
as 
\begin{equation}
D_{\text{DL1}}^{\text{b}}=\ln\left(\frac{p_{\text{b}}}{f_{\text{c}}\cdot p_{\text{c}}+(1-f_{\text{c}})\cdot p_{\text{light}}}\right),\label{eq:DL1b}
\end{equation}
where $p_{F}$ represents the NN output score corresponding to flavor
$F$, and $f_{\text{c}}$ represents the effective c jet fraction
in the background training sample. This allows flexibility in choosing
the c jet fraction in the background \emph{a posteriori} and allows
optimization of the performance after training. This avoids multiple
trainings of the algorithm with varying c jet fractions as was the
case for MV2c00 and MV2c20 variants (cf. Sec. \ref{subsec:First-generation:-BDTs}).
The value of $f_{\text{c}}$, for example, may be set to 8\% in physics
analyses after optimization. The DL1 algorithm improves the light-flavor
and c jet rejections by 30\% and 10\%, respectively, compared to MV2,
at a b jet efficiency of 70\%. This improvement can be attributed
to the use of JetFitter c tagging variables and a deeper NN capable
of effectively exploiting a larger number of inputs.

During Run-2, the ATLAS experiment also developed a track-based RNN
tagger for flavor tagging. This tagger, named RNNIP \citep{ATL-PHYS-PUB-2017-003},
is designed to efficiently leverage intrinsic correlations between
the IPs (and other observables) of the several tracks in a jet, as
opposed to likelihood-based IP taggers, which assume properties of
each track in a jet are independent of all other track. Owing to a
variable number of tracks per jet, a representation in form of variable-length
sequences of particles and an RNN-like architecture resulted to be
the natural choice for this kind of data. A set of basic $\pt$ and
IP selection criteria are applied while selecting tracks that are
fed as input to the RNNIP tagger. As RNNs can handle sequences of
arbitrary lengths, no limit on the maximum number of tracks per jet
is necessary, but a limit of 15 tracks per jet was imposed for ease
of training. The tagger takes as input the transverse and longitudinal
IPs of each selected track, the $\pt$ fraction of the jet carried
by each selected track, the angular distances between the selected
tracks and the jet-axis, and the hit multiplicities of the selected
tracks. Tracks in each jet are ordered by their transverse IP in the
input sequence. This arbitrary-length sequence is then fed into an
LSTM which transforms it to a 50 dimensional vector in the latent
space. This vector is then processed by a dense NN with four output
nodes, corresponding to probabilities of the jet being a b, c, light-flavor,
or hadronic $\tau$ jet, respectively. The RNNIP algorithm improves
light-flavor and c jet rejection by factor of 2.5 and 1.2 times, respectively,
compared to likelihood-based IP taggers, at a b jet efficiency of
70\%.

Again during Run-2, the ATLAS experiment introduced an additional
variant of the DL1 algorithms by additionally incorporating the outputs
of the RNNIP tagger as inputs to DL1. This variant of DL1, dubbed
DL1r \citep{ATLAS:2019fgd,ATLAS:2022qxm}, is thus able to leverage
additional information that is not strongly correlated with the inputs
of the baseline DL1. The number of inputs to DL1r increases to 31,
and it retains the 8-hidden-layer architecture of DL1, but uses a
larger number of nodes per layer, ranging from 256 in the first hidden
layer to 6 in the final. The output layer similarly has three nodes
corresponding to probabilities of the jet being a b, c, or light-flavor
jet. Similar to $D_{\text{DL1}}^{\text{b}}$, a c tagging score for
DL1r can be defined as
\begin{equation}
D_{\text{DL1r}}^{\text{c}}=\ln\left(\frac{p_{\text{c}}}{f_{\text{b}}\cdot p_{\text{b}}+(1-f_{\text{b}})\cdot p_{\text{light}}}\right),\label{eq:DL1rc}
\end{equation}
where $f_{\text{b}}$ represents the effective b fraction in the background
training sample and can be tuned to optimize b jet rejection rate.
The DL1r tagger improves the light-jet and c jet rejection rates by
about 20\% and 12\% respectively, when compared with the DL1 algorithm,
at a b jet efficiency of 70\%. 

The DeepCSV, DL1(r), and RNNIP algorithms are limited by the fact
that only a small number of charged tracks that pass certain physics-motivated
quality criteria are used as inputs to the network. Information about
neutral jet constituents and additional charged tracks that are deemed
less important (such as ones with lower IP or IP significance), are
essentially lost. This led to the development of DeepJet \citep{Bols:2020bkb},
a hybrid NN that utilizes the full information from a significantly
larger number of charged particles, neutral particles, SVs, and global
jet- and event-level variables, resulting in a total of up to 650
input variables. Sixteen features of up to 25 charged particles are
used as input. These features include track kinematics, track fit
quality, IP, and IP uncertainty. Additionally, 6 features of up to
25 neutral particles are provided as input. Twelve features of up
to 4 reconstructed SVs in the jet are also used. These features include
the kinematics of the SV, SV mass, fit quality, and flight distance.
Finally, the global variables include the overall jet kinematics,
particle and SV multiplicities in the jet, and number of PVs (\textit{intime
pileup}\footnote{Pileup refers to overlapping events from additional parton-parton
interactions occurring simultaneously within the same bunch crossing.}) in the event. The DeepJet algorithm (also referred to as DeepFlavour
during early developments \citep{CMS-DP-2017-013}) was practically
implemented in the CMS experiment during Run-2 \citep{CMS-DP-2018-058}.

The architecture of DeepJet is tailored to handle the large input
dimension. For this purpose, separate convolutional filters \citep{5265772}
are trained for each of the three branches (charged, neutral particles,
and SVs). For each branch, a $1\times1$ filter size is chosen to
ensure that all particles in the branch undergo the same feature transformation
without interacting with other particles. This step is aimed at reducing
the dimensionality of the input space, and is in contrast to using
convolutional filters on image grids, where the main aim is to find
clusters of neighboring pixels that share similar features relevant
to object detection. To this end, several layers with decreasing number
of filters are used so that features of each constituent are projected
to a lower dimensional space. Outputs from each convolutional branches
are fed into individual LSTMs. A jet is treated and processed by the
LSTM as a sequence of constituents ordered by their importance. The
outputs of the three LSTMs, as well as the global jet/event features
are concatenated and passed as inputs to a dense NN consisting of
8 layers. The output layer consists of 6 nodes, representing b, bb,
lepb, c, uds, and g. As opposed to DeepCSV, DeepJet splits the b node
into b and lepb, the latter marking b jets containing soft leptons.
It also splits the udsg node into uds and g, making quark-versus-gluon
discrimination possible. On the other hand, the cc node present in
DeepCSV is deprecated and merged with the c node in DeepJet.

The DeepJet algorithm in CMS achieves a 20\% improvement in b tagging
efficiency at a light-jet mistag rate of 0.1\%, when compared with
the DeepCSV algorithm. The improved performance can be attributed
to the use of up to an order of magnitude larger number of inputs,
without applying explicit quality requirements on the input constituents,
and the use of neutral candidates as inputs. This change marks a shift
from using a handful of high-level physics-motivated variables to
using a large number of low-level variables without explicit selections,
and relying on the network to automatically leverage relevant information
from inputs that may contain additional inconsequential and/or correlated
information. More detailed performance comparisons for b and c tagging
using CMS collision data can be found in Refs. \citep{CMS-DP-2018-033,CMS-DP-2018-058,CMS:2021scf,CMS-DP-2021-004,CMS-DP-2023-005,CMS-DP-2023-006,CMS-DP-2023-012}.

\subsubsection{Third generation: Sets and clouds\label{subsec:Third-generation:-Sets}}

The CNN- and RNN-based approaches discussed so far treat jets either
as detector images or as ordered sequences of particles. Ordering
jet constituents using a certain parameter, e.g. the IP significance
or $\pt$, is a necessity in all these use cases. As final-state particles
in jets do not have a natural order, any order that is imposed artificially
may prove to be detrimental to the data representation and hence the
model's performance. A more natural way of representing jets is in
the form of point clouds without assigning any order to the particles.
Deep Set-based approaches (cf. Sec. \ref{subsec:DeepSets}), that
treat jet constituents as permutation invariant, are therefore a more
natural choice to handle particle cloud-based representations of jets.

The ATLAS experiment introduced the Deep Impact Parameter Sets (DIPS)
tagger \citep{ATL-PHYS-PUB-2020-014} after the LHC Run-2 data-taking
period. DIPS is based on the Deep Set architecture, and in particular,
the application of the Deep Set formalism in HEP, referred to as Energy
Flow Networks \citep{Komiske:2018cqr}. Similar to RNNIP (cf. Sec.
\ref{subsec:Second-generation:-Early}), DIPS uses track information
as input but treats them as an unordered, variable-sized set, instead
of explicitly assigning an order. The DIPS architecture applies a
NN to each track, sums over the tracks, and then processes the summed
representation using a dense NN with 2 hidden layers and 100 nodes
per layer. The output layer consists of 3 nodes, corresponding to
the probabilities of the jet being a b, c, or light-flavor jet, respectively.
The summation operation over all tracks in a jet being a permutation
invariant operation, makes the network agnostic to the ordering of
the tracks in the input. On the other hand, the final dense NN accounts
for correlations between tracks. 

The permutation invariant approach results in DIPS having fewer trainable
parameters compared to RNNIP (for similar number of inputs and similar
performance). The operation of processing the tracks can also be parallelized
as opposed to the operation of iteratively processing sequence-like
inputs in RNNs. As a result, DIPS exhibits a training speed per epoch
approximately three times faster than RNNIP, and its inference (prediction)
speed surpasses that of RNNIP by a factor of four, when handling the
same number of input features. In terms of performance, the baseline
version of DIPS having the same input features as RNNIP achieves about
15\% higher light-flavor jet rejection and about 3\% higher c jet
rejection at a b jet efficiency of 70\%. Additionally, the lower complexity
and higher training speed of the DIPS tagger facilitates faster optimization
studies and allows the incorporation of a higher number of tracks
and more features per track. When the track selection is relaxed to
additionally incorporate low-$\pt$ and high IP tracks, and additional
IP features per track are added as inputs, the DIPS tagger improves
light-flavor jet rejection by a factor of \textasciitilde 1.9 and
c jet rejection by a factor of \textasciitilde 1.3 at a b jet efficiency
of 70\%, when compared to the baseline version. These improvements
highlight the importance of physically-motivated ML network architectures
in the task of jet flavor tagging.

The introduction of the DIPS algorithm in ATLAS subsequently resulted
in the introduction of a third variant of the DL1 algorithm, dubbed
DL1d \citep{ATL-PHYS-PUB-2022-047,FTAG-2023-01}, superseding the
DL1r algorithm. Instead of using outputs of the RNN-based RNNIP algorithm
as inputs, DL1d utilizes the outputs of Deep Set-based DIPS as input.
At a b jet efficiency of 70\%, DL1d achieves about 25\% higher light-flavor
jet and 20\% higher c jet rejection compared to DL1r. This is a direct
result of the improved performance of the optimized DIPS tagger compared
to the RNNIP algorithm.

The next improvement in tagging performance in ATLAS was achieved
with the use of GNNs (cf. Sec.\ref{subsec:Graph-Neural-Networks}).
The GN1 algorithm \citep{ATL-PHYS-PUB-2022-027}, introduced during
the beginning of the LHC Run-3 data-taking period, directly leverages
information from tracks inside a jet to predict the flavor of the
jet. This is in contrast to other high-level taggers (like the MV2
and DL1 series) used in ATLAS so far that utilized outputs of intermediate
low-level taggers as inputs and performed the task in two steps. The
GN1 algorithm takes as input the jet $\pt$ and pseudorapidity, and
a variable number of tracks contained inside the jet, with 21 tracking
related variables for each track. The tracking related variables include
several kinematic features of the track with uncertainties, IPs with
their significances, and the number of hits in various tracker layers.
For jets with more than 40 tracks, only 40 tracks with the highest
transverse IP significance are chosen. An additional variant, named
GN1 Lep, utilizes the lepton ID of tracks to indicate whether a track
is reconstructed as an electron, or a muon, or neither.

In addition to flavor tagging, the GN1 is assigned two auxiliary objectives
using training targets based on ``truth'' information available
from simulation, similar to ``truth'' flavors obtained from simulation
and used in flavor tagging. The first objective is to label each track
by its origin, i.e. predict whether a track originates from pileup,
the primary collision, a b hadron, a c hadron originating from a b
hadron decay, a c hadron originating from the primary collision, other
secondary interactions, or is an incorrectly-reconstructed track.
The second objective is to predict whether any given pair of tracks
originates from the same point in space (vertex) \citep{Shlomi:2020ufi},
with a binary prediction as an additional GN1 output. These auxiliary
tasks help in the main task of flavor tagging by acting as a form
of supervised attention \citep{2020arXiv200708294H}, since detecting
tracks from b and c hadrons and being able to group multiple tracks
into a single SV lets the model pay more attention to these tracks
to make flavor predictions. Thus these objectives guide the network
to learn representations of the jet connected to the underlying physics.
The second auxiliary objective also removes the need for dedicated
SV inputs to the model, and also facilitates reconstructing SVs from
the predictions of the network without using standalone vertexing
algorithms.

The GN1 architecture is based on a previous implementation of GNN-based
jet tagger \citep{2020arXiv200208772S}. Each input track with 21
track features (concatenated with the two global jet features) are
fed into a per-track initialization network with three hidden layers,
each containing 64 neurons, and an output layer with 64 nodes, similar
to the DIPS approach. However, unlike DIPS, no summation operation
on the output representation is applied. Instead, a fully-connected
graph is built from the outputs of this network \citep{2021arXiv210514491B},
where each node corresponds to a latent feature vector representing
a track, and each node in the graph is connected to every other node.
Each layer of the GNN aggregates the properties of each input node
and its connections to compute a corresponding output node. This is
achieved in each layer by first processing the input features using
a fully-connected layer, then computing edge scores for each node
pair in the output of the fully-connected layer using a second fully-connected
layer, then computing pairwise attention weights using the edge scores,
and finally taking the weighted sum of each node from the output of
the first fully-connected layer weighted by the respective attention
weights. Three such layers are stacked to construct the GNN. A global
representation of the jet is formed by combining the output representations
of individual tracks through a weighted sum, the weights for which
are learned during training. Subsequently, three distinct fully connected
feedforward NNs operate independently to achieve the various classification
objectives of GN1. A graph classification network using the entire
jet representation predicts the jet flavor, a node classification
network using features from individual nodes representing tracks predicts
track origin, and an edge classification network utilizing features
of edges representing pairs of tracks predict the vertex compatibility
of every pair of tracks.

The GN1 tagger improves c jet rejection by a factor of \textasciitilde 2.1
and light-flavor jet rejection by a factor of \textasciitilde 1.8
compared to DL1r, at a b jet efficiency of 70\%. For the GN1 Lep variant
the respective improvements are by factors of \textasciitilde 2.8
and \textasciitilde 2.5 compared to DL1r, demonstrating the additional
jet flavor discrimination information available from the leptonID
track input. On the other hand, removing the auxiliary objectives
from the GN1 tagger, results in a performance similar to that of the
DL1r algorithm, suggesting that relevant physically-motivated auxiliary
tasks help the network learn the underlying physics relevant to the
task of flavor tagging.

The advent of the transformer architecture and attention mechanism
led to the development of ParticleTransformer (ParT) \citep{Qu:2022mxj}.
Unlike typical implementations of the transformer architecture for
NLP and related tasks, the ParT architecture does not assign positional
encodings to particles in the jet; this aligns with the concept of
particle clouds and keeps the jet constituents permutation invariant.
Instead, the novelty of ParT lies in the use of two sets of inputs:
the ``particle'' input comprising the features associated with all
particles in a jet, and the ``interaction'' input which encodes
pairwise features for every possible pair of particles in the jet.
The latter is, therefore, represented by a $N\times N$ matrix of
values for each pairwise feature, where $N$ is the number of constituents
in a jet. The network can also be viewed as a fully-connected GNN
(cf. Sec. \ref{subsec:Graph-Neural-Networks}) where each node represents
a particle and each edge represents a pairwise feature corresponding
to the two nodes it connects.

The particle inputs and the interaction inputs are projected individually
into latent embeddings using an MLP for each. The particle embedding
is then operated on by a stack of layers implementing MHA (cf. Sec.
\ref{subsec:Transformers}), referred to as particle attention blocks
(PABs). The introduction of attention mechanisms allows the model
to focus on specific aspects of the particle embeddings, enhancing
the network's ability to capture relevant information from individual
particles. At every PAB layer, the same embedded interaction matrix
is used as a weight to augment the attention weights. This mechanism
is crucial for incorporating information from pairwise interaction
data into the particle embeddings, enabling the network to leverage
the interplay between particles and interactions within a jet. The
output from the last PAB, along with a global class token, is fed
into two stacked Class Attention Blocks (CABs), which compute the
attention score between the class token and all the particles. This
allows the model to weigh the importance of different particles with
respect to the class token, and selectively consider information from
various particles when making a classification decision for the entire
jet. The output from the second CAB is then passed to an MLP to produce
the final classification score. This final step consolidates the information
processed through the attention mechanisms and MLPs, in order to perform
the classification.

In Ref. \citep{Qu:2022mxj}, the authors observe that the ParT model
benefits from using a large dataset (JetClass) comprising \textasciitilde 100M
jets, and that other transformer-based jet taggers fail to surpass
the DGCNN-based ParticleNet (cf. Sec. \ref{subsec:Third-generation-boosted})
due to insufficient number of jets in the training samples. This indicates
that models based on transformers are effective at harnessing larger
training datasets by utilizing the attention mechanism.

A version of the ParT architecture was implemented in CMS \citep{CMS-DP-2022-050}
in early Run-3 using three PABs and one CAB, with 8 attention heads
per block. The ParT-based b tagger achieves about 1\% (12\%) higher
b jet efficiency at a light-flavor (c) jet mistag rate of 1\%, when
compared to DeepJet-based b tagger. This gain is about 15\% (35\%)
for jets with $\pt>300$ GeV at same mistag values. Similarly, the
ParT-based c tagger improves c jet efficiency by 14\% (20\%) compared
to DeepJet at a light-flavor (b) jet mistag rate of 1\% for an inclusive
selection of jets, and almost 60\% (60\%) for jets with $\pt>300$
GeV. Detailed performance comparisons can be found in Ref. \citep{CMS-DP-2022-050}.
These indicate that the ParT architecture results in substantial gains
for high-$\pt$ jets, which are usually underrepresented in tagger
trainings.

In the ATLAS experiment, the flavor tagging performance is further
improved using the transformer architecture to supersede the GN1 tagger
with the GN2 tagger \citep{Duperrin:2023elp}. The attention type
was been changed to \emph{ScaledDotProduct} \citep{NIPS2017_3f5ee243}.
Although the change in attention type has no impact on physics performance,
it reduces training time and memory usage. In GN2, the computation
of attention weights is decoupled from the calculation of updated
node representations, incorporating a dense layer between the attention
layers. This introduces additional flexibility and complexity into
the model and allows the model to gain the capacity to learn more
intricate relationships and dependencies within the graph. The learning
rate in GN2 is based on the \emph{One-Cycle} learning rate scheduler
\citep{2018arXiv180309820S}, that adjusts the learning rate during
training, allowing for a faster convergence during the initial phase
and a finer exploration of the loss landscape towards the end. The
number of GNN layers is enhanced to 6, compared to 3 in GN1, enabling
GN2 to capture more complex hierarchical patterns and dependencies
within the graph. The number of learnable parameters in GN2 is 1.5M,
which is almost twice of that in GN1. The number of attention heads
is also enhanced to 8 and the training dataset size was expanded from
\textasciitilde 30 million jets to \textasciitilde 192 million jets.
These updates resulted in an improvement in the light-flavor (c) jet
rejection by a factor of 2 (1.5) compared to GN1, at a b jet efficiency
of 70\% \citep{FTAG-2023-01}, once again demonstrating the impact
of enhanced attention mechanisms and more complex network architectures
capable of leveraging larger training datasets.

\subsection{Multi-pronged jet tagging}

Lorentz-boosted heavy particles decaying into heavy flavor quarks
are usually reconstructed with one large-radius jet. The CMS experiment
used AK jets with a distance parameter of 0.8 (AK8) and 1.5 (AK15)
in Run-2, while the ATLAS experiment used a distance parameter of
1.0 (AK10). In Run-1, jets clustered with the Cambridge/Aachen jet
clustering algorithm \citep{Dokshitzer:1997in} with $\Delta R=0.8$
(CA8) and $\Delta R=1.5$ (CA15) were also used in CMS \citep{CMS-PAS-BTV-13-001}.

\subsubsection{First generation: BDTs and shallow NNs}

In early Run-2 (late Run-1), the CMS experiment demonstrated the effect
of using the same CSVv2 (CSV) algorithm (cf. Sec. \ref{subsec:First-generation:-BDTs})
developed for thin jets to run b tagging on AK8 jets \citep{Sirunyan:2017ezt}
(CA8 and CA15 jets \citep{CMS-PAS-BTV-13-001,CMS-DP-2014-031}). This
was achieved by using a set of looser track-to-jet and vertex-to-jet
requirements for the constituents of an AK8 jet in Run-2, in order
to account for all tracks and vertices within the effective radius
of the jet \citep{Sirunyan:2017ezt}. A second approach is to apply
the CSVv2 algorithm on the subjets of the AK8 jets, where the subjets
are obtained by undoing the last step of the AK algorithm. A yet another
approach is to match AK4 jets with AK8 jets by requiring the angular
separation between the two jet axes to be small (e.g. $<0.4$). In
this case, an AK8 jet is considered tagged based on the CSVv2 score
of at least one of the AK4 jets matched to the AK8 jet.

The tagging performance of these strategies are dependent on the type
of signal and backgrounds being considered and the Lorentz boost of
the originating particle. In the case of tagging a Higgs (H) boson
decaying into a pair of b quarks (H$\rightarrow$$\text{b}\bar{\text{b}}$)
against a gluon decaying to a pair of b quarks (g$\rightarrow$$\text{b}\bar{\text{b}}$),
the subjet tagging strategy achieves about 70\% higher signal tagging
efficiency (\textasciitilde 25\% against \textasciitilde 15\%),
compared to the strategy of tagging the entire AK8 jet, at jet $\pt$
between 300 and 500 GeV and a background mistag rate of 10\%. Tagging
the AK8 jet outperforms tagging matched AK4 jets at these $\pt$ ranges.
On the other hand, at jet $\pt$ above 1200, the subjet tagging method
still outperforms the other two approaches, but the matched-AK4 tagging
approach outperforms the AK8 tagging approach. Detailed performance
comparisons can be found in Ref. \citep{Sirunyan:2017ezt}.

The first dedicated tagger for di-pronged jets in CMS was also developed
in Run-2 and was named \emph{double-b} \citep{CMS-PAS-BTV-15-002,Sirunyan:2017ezt}.
The double-b algorithm was developed to exploit not only the presence
of two b hadrons inside the AK8 jet but also the correlation between
the directions of the momenta of the two b hadron. Besides using the
same variables as CSVv2, the double-b algorithm used track and SV
kinematics computed with respect to the \emph{jet subjettiness} ($\tau$)
axes \citep{Thaler_2011,Thaler:2011gf}. Furthermore, tracks from
all the SVs associated with a given $\tau$ axis are combined to define
the SV mass and $\pt$ corresponding to that axis. A total of 27 jet
features are then combined using a BDT that is trained to discriminate
H$\rightarrow$$\text{b}\bar{\text{b}}$jets from jets in multijet
events. The double-b algorithm slightly exceeds the performance of
subjet tagging at intermediate boosts but improves bb jet tagging
efficiency by about a factor of 2 (30\% against 15\%) at high boosts
($\pt>1.2$ TeV), at a background mistag rate of 10\%. Thus the double-b
algorithm brings a substantial improvement to searches for heavy resonances
that are expected to have high-$\pt$ jets in the final state.

Similar to CMS, the ATLAS experiment used a combination of AK10 jets
and small-radius AK jets with $\Delta R=$ 0.2 or 0.3 to identify
boosted heavy particles in Run-1 \citep{ATLAS:2015zug} and early
Run-2 \citep{ATL-PHYS-PUB-2015-035,ATLAS-CONF-2016-039,ATLAS:2019lwq}.
The AK10 jet reconstructed from calorimeter information was used to
reconstruct the decay, while b tagging algorithms were run on the
individual small-radius track jets\footnote{\emph{Track jets} are reconstructed by applying jet clustering on
the tracks detected in the inner tracker. These jets provide information
complementary to that obtained from jets reconstructed with calorimeter
information. More details can be found in Ref. \citep{ATL-PHYS-PUB-2014-013}.} matched to the AK10 jet. The MV2 algorithm evaluated on each thin
jet, along with requirements on jet substructure variables associated
with the AK10 jet, was used to tag these decays.

In late Run-2, ATLAS made use of variable-radius (VR) track jets associated
with AK10 jets \citep{ATL-PHYS-PUB-2020-019} for di-pronged b tagging.
VR jets are clustered using the VR AK algorithm \citep{Krohn:2009aa,ATL-PHYS-PUB-2017-010}
that uses a varying value of $\Delta R$ depending on the jet $\pt$
and a constant parameter $\rho$, such that the radius scales as $\rho/\pt$.
As this discussion involves the use of both BDT- and RNN-based taggers,
it is again followed up in the next subsection.

\subsubsection{Second generation: Early DNNs}

The late Run-2 approach of the ATLAS experiment towards di-pronged
b tagging evolved with the use of VR track jets associated with an
AK10 jet. Reference \citep{ATLAS:FTAG-2019-006} introduces three
new approaches; the first two involve tagging the individual VR track
jets with the BDT-based MV2 and the RNN-based DL1r taggers, respectively,
the latter being optimized dedicatedly for VR jets \citep{ATLAS:FTAG-2019-006}.
The third approach \citep{ATL-PHYS-PUB-2020-019,ATL-PHYS-PUB-2021-035}
involves using a feedforward DNN to combine the outputs of the single-pronged
tagging algorithms along with AK10 jet kinematics to predict the probability
of the AK10 jet containing the entire heavy particle decay. The inputs
to the NN include kinematics of the fat jet, and output nodes of the
DL1r (or MV2) algorithm (probability of the jet being b, c, or light-flavor)
run on up to three VR jets associated with the AK10 jet. The network
architecture consists of six fully-connected layers, each with 250
nodes. Since DNNs require fixed-length inputs, inputs corresponding
to any missing subjets are replaced with the mean input values when
fewer than 3 subjets are associated with the AK10 jet. In this paper,
the third tagger is denoted by $D_{\text{Xbb}}$.

The baseline DL1r-based VR subjet tagging approach improved the rejection
of multijet (top) background by a factor of \textasciitilde 1.8 (\textasciitilde 1.2)
compared to the MV2-based VR tagging approach, at a Higgs efficiency
of 70\%. At the same signal efficiency, the $D_{\text{Xbb}}$ tagger
has a similar rejection for multijet background, but improves top
rejection by a factor of \textasciitilde 1.9. On the other hand,
at high efficiencies of the Higgs signal (\textasciitilde 93\%),
the $D_{\text{Xbb}}$ tagger rejects about 9 (1.6) times more multijet
(top) background as compared to the VR-based MV2 tagger, while the
DL1r performance remains similar to that at low Higgs boson efficiencies.
As neither the mass of the AK10 jet, nor any variables largely correlated
with the mass, is used as input to $D_{\text{Xbb}}$, the multijet
background jets mistagged by the $D_{\text{Xbb}}$ algorithm do not
exhibit an artificial mass peak. This is an important feature of the
tagger as is discussed in the following paragraphs.

In Run-2, the CMS experiment developed the DeepAK8 algorithm \citep{CMS:2020poo}.
This multiclass classifier was designed to identify hadronically-decaying
heavy particles, with 5 main classes in the output node corresponding
to W, Z, and H bosons, top (t) quark, and other decays. Furthermore,
each main category is subdivided into minor categories depending on
the decay of each particle into a pair of b, c, or light-flavor quarks.
Thus the DeepAK8 algorithm is also able to perform di-pronged flavor
tagging.

The DeepAK8 algorithm uses two lists of inputs corresponding to constituent
particles and SVs, respectively. The ``particle'' list includes
42 variables per particle, including $\pt$, energy, charge, angular
separation between the particle and the jet axis or the subjet axes,
and displacement and quality of the tracks. The ``SV'' list includes
15 features of SVs including the kinematics, displacement, and quality
criteria. The ``particle'' and ``SV'' lists are sorted (in a descending
order) by the particle $\pt$ and the two-dimensional IP significance,
respectively, and the features of only the first 100 particles and
first 7 SVs are used.

As is the case for DeepJet (cf. Sec. \ref{subsec:Second-generation:-Early})
the large input feature space is first processed with two separate
one-dimensional CNNs corresponding to the ``particle'' and ``SV''
lists, respectively. The CNN structure is based on the ResNet model
\citep{2015arXiv151203385H} and is aimed at reducing the dimensionality
and keeping only the relevant information in the intermediate latent
space. The CNN for the ``particle'' (``SV'') list has 14 (10)
layers. The output of the CNNs are fed into a single-layer fully-connected
layer with 512 nodes. This is followed by the output layer with nodes
corresponding to the several particle and decay categories.

Even though the mass of the jet is not explicitly used as input to
DeepAK8, the mass can be inferred by the network from the kinematics
of the particle inputs. Hence, the DeepAK8 network is able to leverage
the difference in invariant masses of the W/Z/H/t particles to make
its predictions. This is, however, an undesirable feature in several
physics analyses as analyses often rely on leveraging mass peaks corresponding
to heavy resonances in a continuum of background. The baseline DeepAK8
tagger preferentially selects jets with masses close to masses of
the signal particles and thereby \emph{sculpts} the background jet
mass spectrum to artificially peak at these masses. This phenomenon,
referred to as \emph{mass sculpting}, is detrimental to the next steps
of physics analyses that may involve using the mass of jets to differentiate
between the signal and background templates. Therefore, a second version
of the DeepAK8 algorithm, dubbed DeepAK8-MD (``mass-decorrelated''),
is trained not to utilize the mass of the jet in performing classification.
In addition to the baseline version, the DeepAK8-MD variant includes
a mass prediction network, consisting of 3 fully-connected layers,
each with 256 nodes. This network is trained to predict the mass of
a background jet from the features extracted by the CNNs. The accuracy
of the mass prediction for the background jets is then added as a
penalty term to the loss function, thereby preventing the CNNs from
learning features that are correlated with the mass.

The baseline version of the DeepAK8 algorithm improves the efficiency
of selecting H$\rightarrow$$\text{b}\bar{\text{b}}$ jets by a factor
of \textasciitilde 1.7 compared to the double-b tagger, at a background
mistag rate of 0.1\% and intermediate boosts of the jets. The DeepAK8-MD
tagger has a performance intermediate to double-b and DeepAK8, as
it is prevented from using the mass of the jet to discriminate signal
from background. The improvement brought by the baseline version is
by a factor of \textasciitilde 2.2 at high $\pt$ (> 1000 GeV), compared
to double-b.

Run-2 of CMS also saw the development of the DeepDoubleX series of
taggers \citep{CMS-DP-2018-046,CMS-DP-2022-041}. These are three
different binary classifiers that aim to classify decays of the Higgs
boson into a pair of b quarks, c quarks, or light-flavor quarks. The
three versions called BvsL, CvsL, and CvsB, as the names suggest,
perform binary classification of AK8 jets to distinguish b from light-flavor,
c from light-flavor, and c from b jets, respectively. Similar to DeepJet,
DeepDoubleX uses properties of charged particles, neutral particles,
and SVs associated with the jet. After pruning, the first 21, 8, and
7 features, respectively, from each collection that provide the most
relevant information, are kept. Additionally, the 5 most relevant
features among the inputs to the double-b algorithm are also kept.
For each of the charged, neutral, and SV collections, separate $1\times1$
convolutions, each with 2 hidden layers with 32 filters each, are
trained. Each of the outputs from the CNNs is then separately is then
fed into a GRU with 50 output nodes. The outputs of the GRUs, along
with the global jet properties, are then processed by a dense layer
with 100 nodes. The output layer then consists of two nodes, which
represent either BvsL, CvsL, or CvsB, depending on the variant.

While earlier versions of the DeepDoubleX models, labelled ``v0''
\citep{CMS-DP-2018-046}, used a similar mass decorrelation technique
as DeepAK8-MD, later versions achieve mass-decorrelation more ``natively''
by using simulated samples of Higgs-like particles with variable masses.
This makes the DeepDoubleX taggers suitable for tagging any heavy
particle resonance decaying into a pair of b or c quarks. The BvsL
tagger improves bb tagging efficiency by a factor of 2 compared to
double-b at a multijet mistag rate of 1\%. The CvsL (CvsB) tagger
achieves a cc tagging efficiency of \textasciitilde 0.48 (\textasciitilde 0.38)
at a multijet (bb) mistag rate of 1\%.

\subsubsection{Third generation: Sets and clouds\label{subsec:Third-generation-boosted}}

The ATLAS collaboration developed the DeXTer (Deep set X$\rightarrow$bb
Tagger) \citep{ATL-PHYS-PUB-2022-042} after the Run-2 data-taking
period, inspired by the architecture of the Deep Set-based DIPS tagger
for single-pronged jets. The DeXTer algorithm is tailored towards
tagging low-$\pt$ ($\pt\lesssim200$ GeV) jets that may arise, e.g.,
from low-mass beyond-SM particles that decay into a pair of b quarks
\citep{Curtin:2013fra,Casolino:2015cza,Cepeda:2021rql}. For this
purpose, large-radius track jets are clustered using the AK algorithm
with $\Delta R=0.8$, using an extended list of tracks found around
a small-radius particle-flow \citep{ATLAS:2017ghe} jet (AK4) with
$\pt>20$ GeV, with a similar approach as that adopted in Ref. \citep{ATLAS:2020ahi}.
This ensures that the particles that may escape the 0.4 radius of
the AK4 jet are captured by the large-radius track jet.

The DeXTer algorithm uses two separate feedforward NNs to extract
features from the tracks and SVs associated to the AK8 track jet.
Each of these contain 2 hidden layers with 100 nodes each, and an
output layer with 128 nodes. The inputs to the track (SV) NN are sorted
by the IP (decay length) significance of the tracks (SVs) and the
first 25 (12) tracks (SVs) are used as inputs. Even though an artificial
ordering is imposed at this stage to select the inputs, the NNs are
permutation invariant in the same way as in the DIPS architecture.
The 256 output nodes from the two NNs, combined with features relating
to the jet's kinematics, are used as inputs to a global feedforward
NN with 3 hidden layers. The output layer of this NN consists of three
nodes, representing the probability of the jet containing two b quarks,
one b quark, or only light-flavor quarks.

Deep Set-based implementations of jet taggers, however, process particles
only in a global way. This falls short of the advantage that CNNs
provide---capturing local information in a global context by using
multiple hierarchical layers with increasing levels of abstraction.
On the other hand, DGCNNs overcome this challenge with stackable EdgeConv
operations, which achieve hierarchical representations similar to
CNNs, starting with an initial particle cloud representation (cf.
Sec. \ref{subsec:Dynamic-Graph-Convolutional}).

The ParticleNet tagger \citep{Qu:2019gqs} implements the DGCNN architecture
with jets representations as inputs in order to fully exploit the
potentials of a particle cloud representation. It makes use of the
EdgeConv operation starting with jet constituents expressed with their
associated features and coordinates in the pseudorapidity--azimuth
plane of the detector. For each particle, its features and those of
the $k$ (e.g. $k=16$) nearest neighboring particles are used to
find the transformed ``edge features'' and EdgeConv operation for
the particle. A total of 3 EdgeConv layers are used, such that the
physical coordinates of the jet constituents are used as inputs to
the first layer, while transformed feature vectors generated as outputs
from each layer are used as inputs to the subsequent layer. Following
the EdgeConv blocks, the outputs are processed by a global average
pooling layer and a fully-connected layer to produce the outputs.

The ParticleNet tagger was implemented in CMS in conjunction with
AK8 and AK15 jets after Run-2 \citep{CMS-DP-2022-005}. Similar to
DeepAK8, the CMS implementation of ParticleNet identifies W/Z/H bosons
and classifies them by their decay modes (bb/cc/light-flavor). The
CMS implementation of the ParticleNet tagger uses SV features as inputs,
in addition to particle features.

A second variant of the ParticleNet tagger, named ParticleNet-MD \citep{CMS-DP-2022-005},
is trained with simulated samples containing Higgs-like particles
with a flat mass spectrum between 15--250 GeV. This tagger contains
4 output nodes, corresponding to the probabilities $P($X$\rightarrow$bb$)$,
$P($X$\rightarrow$cc$)$, $P($X$\rightarrow$qq$)$, and $P($QCD$)$.
Deriving from these, flavor tagging discriminants, defined as
\begin{align*}
D_{\text{X\ensuremath{\rightarrow}bb}} & =\frac{P(\text{X\ensuremath{\rightarrow}bb})}{P(\text{X\ensuremath{\rightarrow}bb})+P(\text{QCD})},\\
D_{\text{X\ensuremath{\rightarrow}cc}} & =\frac{P(\text{X\ensuremath{\rightarrow}cc})}{P(\text{X\ensuremath{\rightarrow}cc})+P(\text{QCD})},
\end{align*}
are useful in tagging heavy particles decaying to a pair of b quarks
and a pair of c quarks, respectively, against the massive, irreducible
QCD background usually encountered in physics analyses. ParticleNet-MD
improves H$\rightarrow$$\text{b}\bar{\text{b}}$ tagging efficiency
by \textasciitilde 30\% compared to DeepAK8-MD, at a QCD mistag rate
of 0.1\%. The improvement is around \textasciitilde 50\% for (H$\rightarrow$$\text{c}\bar{\text{c}}$)
tagging. The ParticleNet-MD tagger was also trained for AK15 jets
\citep{Sirunyan:2020aa,VHcc} in CMS. Detailed performance comparison
between double-b, DeepDoubleX, DeepAK8-MD, and ParticleNet-MD in CMS
collision data can be found in Ref. \citep{CMS-PAS-BTV-22-001}. Reference
\citep{CMS-PAS-BTV-22-001} also presents a novel use case of BDTs
in calibration of fat jet flavor tagging algorithms.

Inspired by the transformer- and GNN-based GN2 algorithm for single-pronged
b tagging, the ATLAS experiment developed the GN2X tagger \citep{ATL-PHYS-PUB-2023-021}
for di-pronged b tagging in Run-3. GN2X takes as input 3 variables
($\pt$, pseudorapidity, mass) of the jet, and 20 variables associated
with each track contained in the jet. Up to 100 tracks (with the highest
IP significance) per jet are used as inputs. The architecture is very
similar to that of the GN2 tagger, with a per-track initialization
network designed in the same way as Deep Set formalism without the
reduction (summation) operation over track output features. The outputs
of the initialization networks are fed into a transformer encoder
\citep{2021arXiv211009456S}. A total of 6 encoder blocks with 4 attention
heads are used. The outputs are combined to form a global representation
by computing a weighted sum over the track representations, where
the weights are attention weights learned in the training process.
The global representation is used for classification. The output layer
contains a node representing jets from hadronic top pair samples,
besides the usual H$\rightarrow$bb, H$\rightarrow$cc, and QCD multijet
nodes. In addition, as is the case for GN1 and GN2, two auxiliary
training objectives are also incorporated for GN2X. Track origin classification
and computation of vertex compatibility of any pair of input tracks
are achieved by extending the architecture to include three hidden
layers containing 128, 64, and 32 nodes, respectively, for each task.
Mass-sculpting of tagged background jets is avoided in the same way
as in DeepDoubleX, i.e. by training the networks with signal samples
containing artificial Higgs-like particles with a flat mass spectrum.

The GN2X tagger improves the rejection of top pair (QCD) jet backgrounds
by a factor of \textasciitilde 1.7 (\textasciitilde 1.9) compared
to the $D_{\text{Xbb}}$ algorithm at a H$\rightarrow$bb signal efficiency
of 70\%. Similarly for H$\rightarrow$cc tagging, the top pair (QCD)
jet background rejection is improved by a factor of \textasciitilde 2.6
(\textasciitilde 6.0) at a H$\rightarrow$cc signal efficiency of
70\%. The H$\rightarrow$bb rejection is also improved by a factor
of \textasciitilde 3 at the same H$\rightarrow$cc efficiency, compared
to $D_{\text{Xbb}}$. 

Two additional variants of the GN2X tagger incorporating heterogenous
inputs were also developed. They are referred to as GN2X+Subjets and
GN2X+Flow. The former utilizes the kinematic and b tagging information
of the VR subjets of the AK10 jet, where the subjets are tagged using
the GN2 tagger. The latter variant uses Unified Flow Object (UFO)
\citep{ATLAS:2020gwe} constituents which includes the use of charged
and neutral calorimeter information. GN2X+Subjets improves top jet
rejection by a factor of 2 compared to baseline GN2X, at a H$\rightarrow$bb
signal efficiency of 70\%, but reduces QCD rejection to almost half
of the baseline. On the other hand, the GN2X+Flow architecture improves
the top (QCD) jet rejection by a factor of \textasciitilde 1.6 (\textasciitilde 1.3)
at the same efficiency. 

\begin{sidewaystable*}
\renewcommand{\arraystretch}{1.2}

\begin{centering}
\begin{tabular}{>{\raggedright}m{0.12\textwidth}>{\centering}m{0.1\textwidth}>{\centering}m{0.1\textwidth}>{\centering}p{0.1\textwidth}>{\centering}p{0.1\textwidth}>{\centering}p{0.1\textwidth}>{\raggedright}m{0.25\textwidth}}
\hline 
\textbf{Generation} & \textbf{Main architecture} & \multicolumn{2}{c}{\textbf{Single-pronged (AK4)}} & \multicolumn{2}{c}{\textbf{Multi-pronged}} & \textbf{Novelty}\tabularnewline
 &  & \textbf{CMS} & \textbf{ATLAS} & \textbf{CMS (AK8/AK15)} & \textbf{ATLAS (AK10)} & \tabularnewline
\hline 
\multirow{2}{0.12\textwidth}{\textbf{First} (Run-1 and early Run-2, shallow architectures)} & Shallow NN & CSVv2 & MV1 & CSVv2 & MV1 using subjets & \multirow{2}{0.25\textwidth}{Switch from likelihood methods to first ML-based methods. Incorporate
larger number of inputs.}\tabularnewline
 & BDT & c-tagger, cMVAv2 & MV2 & double-b & MV2 using subjets & \tabularnewline
\hline 
\multirow{4}{0.12\textwidth}{\textbf{Second} (Late Run-2, early deep NNs) } & Deep dense NN & DeepCSV & DL1 &  &  & Use information from multiple selected tracks. Multiclassification.\tabularnewline
 & CNN &  &  & DeepAK8 &  & Use a large number of particles per jet.\tabularnewline
 & RNN &  & RNNIP, DL1r &  & DL1r using subjets, $D_{\text{Xbb}}$ & Accommodate an arbitrary number of tracks per jet.\tabularnewline
 & CNN+RNN & DeepJet &  & DeepDoubleX &  & Avoid applying selection criteria; use all jet constituents of jets
including neutral particles.\tabularnewline
\hline 
\multirow{4}{0.12\textwidth}{\textbf{Third} (Early Run-3, sets and particle cloud represent$\-$ations)} & Deep Set &  & DIPS, DL1d &  & DeXTer & Avoid sorting of jet constituents: permutation invariant.\tabularnewline
 & GNN &  & GN1 &  &  & Use large number of track features, including pixel hits. Assign auxiliary
track origin and vertex finding tasks.\tabularnewline
 & DGCNN &  &  & ParticleNet &  & Leverage local features of point clouds using convolutions.\tabularnewline
 & Transformer & ParticleTrans\-formerAK4 & GN2 &  & GN2X & Leverage pairwise features, attention mechanisms, and larger training
datasets.\tabularnewline
\hline 
\end{tabular}
\par\end{centering}
\caption{\label{tab:Flavor-tagging-algorithms}Flavor tagging algorithms developed
in the CMS and ATLAS experiments, classified by generations and main
ML architecture used.}
\end{sidewaystable*}

\section{Discussions and future prospects\label{sec:Discussions}}

Section \ref{sec:Machine-Learning-in} highlights how ML-based flavor
tagging algorithms have evolved over time in the CMS and ATLAS experiments.
These architectures have not only been utilized for flavor tagging
but also paved the way for identifying hadronically-decaying heavy
objects like the W boson and the top quark. At times, these experimental
innovations have utilized progress from parallel developments in phenomenological
studies. Since this review focuses only on flavor tagging algorithms
that have been successfully used at the LHC, neither generalized heavy-object
tagging nor the parallel developments in phenomenology have been discussed.
However, it is important to note that several of the methods discussed
here are the consequence of earlier research into QCD-motivated jet
substructure methods. These earlier methods proposed and improved
definitions of jet substructure variables that can be used to tag
jets. These observables utilize the internal kinematic properties
of a highly boosted jet to identify W \citep{Cui:2010km,Thaler_2011},
Z/H \citep{Butterworth:2008iy,Plehn:2009rk,Larkoski:2013eya,Moult:2016cvt},
and top \citep{Kaplan:2008ie,Thaler:2011gf,Plehn:2011tg,Plehn:2011sj,Soper:2012pb,Anders:2013oga,Kasieczka:2015jma}
jets, as well as to distinguish quark- and gluon-initiated jets \citep{Gallicchio:2011xq,Gallicchio:2012ez,Larkoski:2014pca,Bhattacherjee:2015psa,FerreiradeLima:2016gcz,Frye:2017yrw}. 

Similarly, ML-based tagging strategies first introduced in phenomenological
studies have also been used in developing modern flavor tagging algorithms
in experiments. As successors of substructure-based approaches, ML-based
studies have interpreted jets as detector images \citep{Cogan:2014oua,Almeida:2015jua,Oliveira:2016aa,Baldi:2016fql,Schwartzman_2016,Komiske:2016rsd,ATL-PHYS-PUB-2017-017,Kasieczka:2017nvn,Macaluso:2018tck,Andrews:2019faz,Diefenbacher:2019ezd,Bhattacharya:2020aid,Andrews:2021ejw}
using CNNs, particle sequences and trees \citep{Guest:2016iqz,Pearkes:2017hku,Egan:2017ojy,Fraser:2018ieu,ATL-PHYS-PUB-2017-003,CMS-DP-2017-013,CMS-DP-2017-049,Butter:2017cot,Louppe:2017ipp,CMS-DP-2018-058,CMS-PAS-JME-18-002,Kasieczka:2018lwf,Erdmann:2018shi,Cheng:2018aa}
using RNNs, sets using energy flow networks \citep{Komiske:2018cqr,PhysRevD.103.074022},
interactions using interaction networks \citep{Moreno:2020aa}, and
graphs and particle clouds \citep{nips,Abdughani:2018wrw,Qu:2019gqs,Roy:2019jae,REN2020135198,Mikuni:2020wpr,Chakraborty:2020yfc,Dreyer:2020brq,Atkinson:2021nlt,Atkinson:2022uzb,Dreyer:2021hhr,Shimmin:2021pkm,Dreyer:2022aa,Li:2022xfc,Gong:2022lye,Qu:2022mxj,Ma:2022bvt}
using GNNs, DGCNNs, autoencoders \citep{JMLR:v11:vincent10a}, and
transformers \citep{NIPS2017_3f5ee243}. More detailed reviews of
some of these algorithms can be found in Refs. \citep{Guest:2018yhq,Kogler:2018hem,Kasieczka:2019dbj,LARKOSKI20201}.
While early developments with ML provided performances at par with
those obtained with substructure observables \citep{Kasieczka:2017nvn,Moore:2018lsr},
more recent developments have significantly surpassed earlier approaches
in terms of performance, often at the cost of interpretability \citep{Romero:2021qlf,Khot:2022aky}.
Other phenomenological studies have proposed using novel jet reconstruction
techniques to reconstruct hadronically-decaying heavy particles and/or
enhance tagging performance \citep{Stewart:2015aa,Thaler:2015aa,refId0,Mukhopadhyaya:2023aa,larkoski2023jet,Ju:2020tbo,Mondal:2023law}
.

Trends in the evolution of tagging algorithms indicate a steady shift
towards leveraging increasingly lower levels of information along
with advanced networks are that capable of extracting relevant features
from a large number of inputs. As leveraging low-level inputs preserves
information that might be lost in the process of constructing physically-meaningful
quantities, future tagging algorithms in the ATLAS and CMS experiments
are expected to directly use raw information from the detectors. On
the other hand, the examples of improvements observed by adding auxiliary
tasks in GN1/GN2 (cf. Sec. \ref{subsec:Third-generation:-Sets}) and
heterogenous inputs in GN2X+Flow (cf. Sec. \ref{subsec:Third-generation-boosted})
algorithms, indicate that using physics-motivated high-level quantities
to steer the learning priorities of the network can bring about additional
improvements in tagging performance. Thus, future taggers are expected
to leverage both detector-level information and high-level targets
simultaneously, with mechanisms to effectively guide the network to
learn physically-meaningful representations. Using detector-level
information, however, requires even more advanced architectures that
can condense and interpret a large number of input variables, along
with faster, parallelizable computing methods, which are challenges
for the broader ML community to address. Moreover, leveraging larger
number of input features usually requires larger number of jets in
the training process, and hence future training datasets may need
$\mathfrak{\mathcal{O}}(10^{9})$ jets which poses significant computing
and storage challenges.

However, with the use of lower-level information and attempts to extract
every last bit of information available in the training sample, there
is a possibility that newer taggers simply learn to leverage simulation-specific
artifacts that do not reflect real collision data. Simulations of
jets are, in turn, limited by the jet showering algorithms' precision
and underlying assumptions, among other factors. Therefore, it is
inevitable that at some point any additional gains observed from new
tagging algorithms are largely due to the tagger backtracing the steps
in the algorithms used to generate the training samples or relying
exceedingly on mismodeled features in simulation \citep{Nachman:2019yfl,Yallup:2022rjd,Butter:2022xyj}.
Such gains in performance in simulation would not necessarily be reflected
when the algorithm is used in conjunction with collision data. Data-aware
and unsupervised training methods \citep{Metodiev:2017vrx,Andreassen:2018apy,Komiske:2018oaa,Alvarez:2022qoz}
and adversarial training approaches \citep{Stein:2022nvf,CMS-DP-2022-049}
have been proposed to mitigate this problem, but may adversely affect
the best achievable performance in simulations. Therefore, while improved
tagging performance on simulations are interesting from an ML point
of view, they may not necessarily be indicative of the achievable
performance in collision data. Recent studies have also used QCD-motivated
observables to design more interpretable networks with fewer learnable
parameters \citep{Komiske:2017aww,Grojean:2020ech,Bogatskiy:2020tje,Munoz:2022gjq,Fedkevych:2022mid,Bhattacherjee:2022gjq,Bradshaw:2022qev,Das:2022cjl,Cal:2022fnm}.
These approaches provide an alternative avenue for developing more
robust taggers without relying exceedingly on features in simulations
that are prone to mismodeling.

The above paragraph motivates the necessity to calibrate flavor tagging
algorithms using collision data. Corrections are usually derived as
Scale Factors (SFs) that measure the ratio of the tagging efficiency
in data to that in simulation \citep{ATLAS-CONF-2018-006,ATLAS-CONF-2018-045,ATLAS-CONF-2018-055,ATLAS:2019bwq,ATL-PHYS-PUB-2021-004,ATLAS:2021cxe,ATL-PHYS-PUB-2022-025,ATLAS:2023lwk,CMS-PAS-BTV-13-001,Sirunyan:2017ezt,CMS:2021scf,CMS-DP-2022-005,CMS-PAS-BTV-22-001,CMS-DP-2023-005,CMS-DP-2023-006}.
Performance measurements of the most recent taggers \citep{ATLAS:2023lwk,ATLAS:2021cxe,CMS-DP-2023-005,CMS-DP-2023-006,CMS-PAS-BTV-22-001}
using collision data have yielded SFs that are reasonably close to
1. Significant departures of the SFs from 1 in future taggers, if
observed, would indicate that improvements seen on simulated samples
are a result of the tagger leveraging simulation-only artifacts. In
such cases, understanding and mitigating the sources of mismodeling
in simulations should take precedence over improving tagger performance
on simulations.

Another potential challenge for future taggers is the high level of
pileup expected at the High Luminosity LHC (HL-LHC) \citep{ZurbanoFernandez:2020cco}.
An increased number of particles unrelated to the fragmentation of
b/c hadrons in a jet will inevitably result in poorer tagging performance
at high pileup environments. Improved implementations of attention-mechanisms
and novel pileup mitigation techniques \citep{Bertolini:2014bba,Komiske:2017ubm,Hansen:2018osj,ArjonaMartinez:2018eah,Alipour-fard:2023yjz}
are expected to mitigate such phenomena. 

\section{Conclusion\label{sec:Conclusion}}

Flavor tagging in the CMS and ATLAS experiments at the LHC has seen
significant evolution over the past decade. Tagging algorithms used
in the experiments have been classified into three generations in
this paper. The first generation of taggers, developed and implemented
during Run-1 and early Run-2 of the LHC, mark the first departure
from physics-motivated likelihood methods to shallow ML techniques.
The second generation of taggers, developed during Run-2 and implemented
on legacy Run-2 datasets, are characterized by the use of deep learning
methods and a use of a larger number of inputs. The third generation,
developed shortly before and during early Run-3, marks the use of
more natural and physically-motivated representations for jets, the
use of even larger number of features of jet constituents, and the
use of attention mechanisms. 

The accuracies of these taggers, in case of both single-pronged and
multi-pronged jets, have improved by leaps and bounds over the three
generations. Notably, they have enabled the first observation of the
Higgs boson decaying to a pair of bottom quarks \citep{ATLAS:2018kot,CMS:2018nsn}
and the Z boson decaying to a pair of charm quarks \citep{VHcc,ggHcc}
at the LHC. They have also enabled development of methods to measure
the Higgs-charm coupling \citep{Aad:2022aa,VHcc} to a degree of accuracy
that was earlier predicted to be achievable only at the HL-LHC \citep{ATLAS:2018tmw,ATLAS:2021hui}.

Future tagging algorithms are expected to incorporate increasingly
lower-level information from the detectors, possibly even raw detector-level
information, in conjunction with new architectures that are capable
of extracting information from a large number of inputs. However,
these approaches may be limited by mismodelings that are inevitable
in simulations. Data-aware training methods are expected to circumvent
this issue. The upcoming decades will undoubtedly witness fascinating
advancements in the field of flavor tagging and the broader field
of machine learning.

\bibliographystyle{lucas_unsrt}
\bibliography{review}

\end{document}